\documentclass[aps,prb,twocolumn,showpacs,groupedaddress]{revtex4}
\usepackage{graphicx}

\begin{document}


\title{History effects and pinning regimes in solid vortex matter}

\author{S. O. Valenzuela}
\altaffiliation[Present address: ]{Physics Department, Harvard
University, Cambridge, MA 02138.} 
\affiliation{Laboratorio de Bajas Temperaturas, Departamento de
F\'{\i}sica, Universidad Nacional de Buenos Aires, Pabell\'on I,
Ciudad Universitaria, 1428 Buenos Aires, Argentina}
\author{V. Bekeris}
\affiliation{Laboratorio de Bajas Temperaturas, Departamento de
F\'{\i}sica, Universidad Nacional de Buenos Aires, Pabell\'on I,
Ciudad Universitaria, 1428 Buenos Aires, Argentina}


\author{}
\affiliation{}


\date{\today}

\begin{abstract}
We propose a phenomenological model that accounts for the history
effects observed in ac susceptibility measurements in
YBa$_2$Cu$_3$O$_{7}$ single crystals [Phys. Rev. Lett.
\textbf{84}, 4200 (2000) and Phys. Rev. Lett. \textbf{86}, 504
(2001)]. Central to the model is the assumption that the
penetrating ac magnetic field modifies the vortex lattice
mobility, trapping different robust dynamical states in different
regions of the sample. We discuss in detail on the response of the
superconductor to an ac magnetic field when the vortex lattice
mobility is not uniform inside the sample. We begin with an
analytical description for a simple geometry (slab) and then we
perform numerical calculations for a strip in a transverse
magnetic field which include relaxation effects. In calculations,
the vortex system is assumed to coexist in different pinning
regimes. The vortex behavior in the regions where the induced
current density $j$ has been always below a given threshold
($j_c^>$) is described by an elastic Campbell-like regime (or a
critical state regime with local high critical current density,
$j_c^>$). When the VS is shaken by symmetrical (e.g. sinusoidal)
ac fields, the critical current density is modified to $j_c^< <
j_c^> $ at regions where vortices have been forced to oscillate by
a current density larger than $j_c^>$. Experimentally, an initial
state with high critical current density ($j_c^>$) can be obtained
by zero field cooling, field cooling (with no applied ac field) or
by shaking the vortex lattice with an asymmetrical (e.g. sawtooth)
field. We compare our calculations with experimental ac
susceptibility results in YBa$_2$Cu$_3$O$_{7}$ single crystals.
\end{abstract}


\maketitle

\section{Introduction}

The competition between vortex-vortex interactions and randomness
determine the electromagnetic response of type-II superconductors
in the mixed state.\cite{blatter94} The interaction between
vortices favors an ordered vortex lattice (VL) that opposes the
disorder arising from random pinning centers and temperature. The
combined effect of these interactions leads to a great variety of
collective behaviors, making the VL a paradigmatic system to study
an elastic object in a disordered environment.\cite{crabtree97}

The rich static phases of vortex matter in high $T_c$
superconductors have drawn a great deal of attention in the
past,\cite{blatter94,brandt95,cohen97} but recently, much effort
has been directed towards understanding vortex lattice dynamics both in low
and high $T_c$ materials. Though details are not completely
understood, consensus has been reached that the nucleation and
annihilation of defects (\textit{e.g.}, dislocations) in the VL
might play a major role in its response to an applied
force.\cite{kupfer77,mullock85,kes83,wordenweber86a,shi91,henderson98,xiao99,banerjee99a,andrei99,ravikumar00,SOVPRL00,SOVPRL01}
Reproducible states with distinct mobilities have been observed,
and it has been argued that this reflects distinct degrees of
topological order of the VL: a defective VL would interact more
efficiently with pinning centers than an ordered one because of a
reduction in its effective shear modulus. In particular, several
experiments have revealed that, in certain conditions, an ac
magnetic field or an ac current can assist the VL in
ordering.\cite{henderson98,xiao99,banerjee99a,andrei99,ravikumar00,SOVPRL00,SOVPRL01,ling01}

At first, two basic mechanisms have been invoked to account for
the dynamic ordering of the VL in low $T_c$ superconductors: a
dynamic transition occurring at current densities above a certain
threshold\cite{bhattacharya93,koshelev94,crabtree97} and an
equilibration process where an ac magnetic field (or current)
perturbation assists the VL in going from a supercooled disordered
state into the more ordered equilibrium state. However, data have
been reported that can not be accounted for within either of these
two frameworks. We refer to the recent findings in the vortex
lattice in twinned YBa$_2$Cu$_3$O$_{7}$ (YBCO) single
crystals.\cite{SOVPRL00,SOVPRL01} It was found that when vortices
are shaken by a temporarily symmetric ac magnetic field
(\textit{e.g.}, sinusoidal), they are driven into a mobile state,
but if the shaking field is temporarily asymmetric (\textit{e.g.},
sawtooth), the lattice is trapped into a more pinned
structure.\cite{SOVPRL01} As suggested in Ref.
\onlinecite{SOVPRL01}, the possibility of switching to a high
mobility VL state by applying a symmetrical (sinusoidal, square,
triangular, etc.) ac magnetic field or to a low mobility VL state
by applying a temporarily asymmetric magnetic field waveform
(sawtooth, etc.) is inconsistent with a naive equilibration
process. In addition, the induced current density that
participates in ordering or disordering the VL can be tuned to be
comparable by adjusting the magnetic field amplitude and
frequency. Disordering observed for higher field sweep rates than
those that are used in symmetric oscillating fields that order the
vortex lattice is a clear indication that the higher mobility of
an ordered VL is not attained through a dynamical transition at
high current densities.\cite{SOVPRL01}  If this were the case, the
VL should also order with the asymmetric ac field that is expected
to induce a higher current density.\cite{com} This demonstrates,
at least for YBCO, that the high or low VL mobility results from a
collective dynamical process of a different kind that we believe
is intrinsic to the oscillatory motion.

The attained degree of mobility is found to persist after
interrupting the ac excitation for long periods of time (exceeding
1 h). As the ac field penetration is dependent on its amplitude,
$H_0$, and temperature, $T$, different regions of the sample may
remain in different pinning regimes depending on the detailed
thermomagnetic history of the sample.\cite{SOVPRL00} The dynamical
rearrangement of the VL occurs in the outer zone of the sample
where the penetration of the ac magnetic field induces ac currents
that force vortices to move. An increment of $H_0$ or $T$ will
result in an increase in the penetration depth of the ac field. As
the ac field further penetrates into the sample, the VL mobility
changes. In Ref. \onlinecite{SOVPRL00}, we argued that, as long as
the current density does not exceed the critical one (so that a
small fraction of the vortices moves out of their pinning sites)
the VL will remain in the low mobility state. In a
bulk-pinning-dominated regime, this will happen in the inner
nonpenetrated portion of the sample. As $T$ or $H_0$ increases
this region will shrink and, eventually, disappear. The robustness
of the states involved and their spatial distribution in the
sample are the main ingredients leading to the observed memory
effects in Ref. \onlinecite{SOVPRL00}.

In this paper, we present analytical and numerical calculations
for the field and  current distributions, as well as for the ac
susceptibility $\chi_{ac}$ in samples with inhomogeneous pinning.
If history effects and flux creep can be neglected, the current
and flux densities inside the superconductor are often well
described by the Bean critical state model\cite{bean64} (CS) which
assumes a critical current density that is magnetic field
independent and a null lower critical field, $H_{c1}$ ($H_{c1}=0$,
$B=\mu_0H$). Exact analytical solutions have been obtained for the
perpendicular
geometry\cite{zhu93,mikheenko93,brandt93,clem94,zeldov94} which is
more appropriate for our experimental arrangement. For more
general situations, as in our particular case, a time-dependent
theory is required. If it is assumed that the critical current
density does not vary over the sample thickness, this dynamic
theory can be formulated in terms of a one-dimensional integral
equation as discussed by E. H. Brandt.\cite{brandt94a,brandt94b}
With the use of an appropriate constitutive relation $E(j)$ it is
possible to compute the flux motion (and the magnetization) in
strips, disks, etc. with homogeneous or inhomogeneous $j_c$ when
the applied field is cycled. From the magnetization curves, we
calculate the susceptibility. These results are compared with
measured ac susceptibility data obtained in a twinned
YBCO crystal following a specific protocol
to control the spatial pattern of the dynamical behavior of the VL
in the sample.

As an example, the VL can be prepared by  zero ac field cooling
(ZF$_{ac}$C), cooling the sample below the critical temperature
before applying the magnetic ac field. This traps the VL in a low
mobility state\cite{SOVPRL00} with a high critical current
density, $\sim j_c^>$. If a sinusoidal ac field is then applied,
the VL will become more mobile in the penetrated outer zone of
the sample and it will be characterized by a lower critical
current density, $\sim j_c^<$. Vortices in the inner region of the
sample, where the induced current density $j$ has been always
below $j_c^>$, will remain in a more pinned structure in which
vortices oscillate confined in their pinning potential wells.

The critical current patterning so defined may be observed, for example,
by removing the ac field and applying an increasing ac probe to
measure the ac susceptibility $\chi_{ac}$. This is the basic
protocol that we have developed in calculations and in our ac
susceptibility measurements to show evidence of the different
local degrees of VL mobility. Analytical calculations that
qualitatively introduce our model are presented in Section
\ref{stat}. In Section \ref{dyn} we present our numerical
calculations of $\chi_{ac}$ that take into account a more adequate
transverse geometry and also include the effect of flux creep. In
calculations, we consider two possible values for the critical
current density though it should be kept in mind that it is
observed experimentally that $j_c$ depends on the number of cycles
of the ``shaking" ac field (see Ref. \onlinecite{SOVPRL01}).
Experimental results are described in Section \ref{exp} and
conclusions are drawn in Section \ref{conc}.

\section{Theory}

For symmetrical shaking fields, we can argue that the VL in the
inner portion of the sample will remain in a low mobility state as
long as the inner current density does not exceed the critical
current density. In this case, a Campbell-type regime may describe
the elastic pinning of vortices that are confined to oscillate
inside a potential
well.\cite{campbell69,campbell71,campbell72,konczykowski93,brandt94c}

The main criterion to discriminate between a Campbell and a
critical state regime arises from comparing the scale of the
vortex displacement, $\Delta x$, produced by the induced currents,
and the size of the pinning potential well, $r_p$. The application
of an ac magnetic field results in a periodic compression of the
VL, with a maximum effective vortex displacement at the sample's
boundary. This displacement can be easily estimated from the
variation of the lattice parameter $a_0
=\sqrt{\frac{2}{\sqrt{3}}\frac{\phi_0}{B_{dc}}}$ that occurs after
applying a perturbation in the dc field, $B_{dc}$. For a sample
with no pinning and with $\mu_0 H = B$ (\textit{i.e.}, neglecting
$H_{c1}$), the vortex displacement at the sample perimeter arising
from an ac field of amplitude $B_0$ ($B_0 << B_{dc}$) is $\Delta x
\approx \frac{1}{2}\frac{B_0}{B_{dc}} R$ where $R$ is the sample
radius. If $\Delta x$ is smaller than $r_p$ the elastic pinning
regime usually describes satisfactorily the penetration of
magnetic flux. As the ac field amplitude increases, the region in
the sample where the elastic regime applies naturally shrinks. We
shall come back to this criterion further on, and momentarily we
shall address the problem of a superconductor with nonuniform
critical current density.

\subsection{\label{stat}Statics. Analytical calculations}

\begin{figure}[t]
\vspace{-10mm}
\includegraphics[width=3.5in]{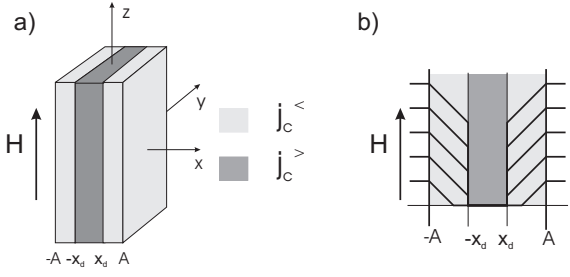}
\vspace{-22mm} \caption{(a) Infinite slab for $\mathbf{\hat{z}}$
and $\mathbf{\hat{y}}$ used in the estimation of the ac magnetic
susceptibility. The field is applied in the $\mathbf{\hat{z}}$
direction. The critical current density is inhomogeneous, with
$j_c = j_c^>$ for $|x|<x_d$ (dark gray) and  $j_c = j_c^<$ for
$|x|>x_d$ (light gray). (b) Field profiles inside the sample for
$j_c^>\rightarrow\infty$. } \label{f:slab}
\end{figure}

We start by calculating  the ac susceptibility of an infinite slab
($| x |  \leq A$) in a longitudinal magnetic field  for which the
inner critical current density,
 $j_c^{int}=j_c^{>}$ ($| x | \leq x_c \leq A$) is larger than
the outer one, $j_c^{ext}=j_c^{<}$ ($ x_c < | x | \leq A$) [see
Fig. \ref{f:slab}(a)]. This first calculation will provide a {\it
qualitative} description of the observed ac field amplitude
dependence of the ac susceptibility. We shall not consider at
present that the ac field modifies the local critical current
density, and we will suppose a given fixed spatial distribution of
$j_c$. For illustrative purposes we also take $j_c^{>} \rightarrow
\infty$ so that the outer region of the sample $x_d <| x | \leq A$
is in the CS whereas there is perfect shielding (Campbell penetration
length $\lambda_C \ll x_d$) for $| x | < x_d$. The calculation of
$\chi_{ac}$ in this simplified situation is straightforward and
follows from known results for the critical state in a slab
\cite{goldfarb91} if one notes that: \textit{i}) for ac field
amplitudes below $H_d = j_c^< (A - x_d)$, $\chi_{ac}$ is the well
known CS susceptibility for an infinite slab with $ j_c =
j_c^{<}$; \textit{ii}) for fields higher than $H_d$, the sample
magnetization is given by the sum of the magnetization due to the
outer region of total width $2(A-x_d)$ in the CS and that due to the
inner region of width $2 x_d$ with perfectly diamagnetic
susceptibility $\chi' - i\chi '' = -1$ [see Fig. \ref{f:slab}(b)].
For an applied ac
magnetic field, $H_{ac} = H_0 \cos(\omega t)$, the susceptibility
is
\\

\noindent $H_{0}\leq H_d:$
\begin{eqnarray}
\label{e:slab1} \chi '&=&-1+\frac{1}{2}x \\
\label{e:slab2}\chi''&=&\frac{2}{3\pi}x\ \label{e:chi1}
\end{eqnarray}

\noindent $H_{0} > H_d:$
\begin{eqnarray}
\chi'=&&\frac{1}{\pi}[(\frac{1}{2}x^*-1)\cos^{-1}(1-\frac{2}{x^*})+
(-1+\frac{4}{3x^*}-\frac{4}{3x^{*2}})\nonumber\\ &&
(x^*-1)^{1/2}]\frac{A-x_d}{A}-\frac{x_d}{A}\
\end{eqnarray}

\begin{equation}
\chi''=\frac{1}{3\pi}(\frac{6}{x^*}-\frac{4}{x^{*2}})(\frac{A-x_d}{A})\label{e:chi2}
\end{equation}

\noindent where $x\equiv \frac{H_0}{H_p}$, $H_p=j_c^< A$ and
$x^*=\frac{H_0}{H_d}=\frac{H_p}{H_d}x=\frac{A}{A-x_d}x$.

\begin{figure}[h]
\includegraphics[width=3.7in]{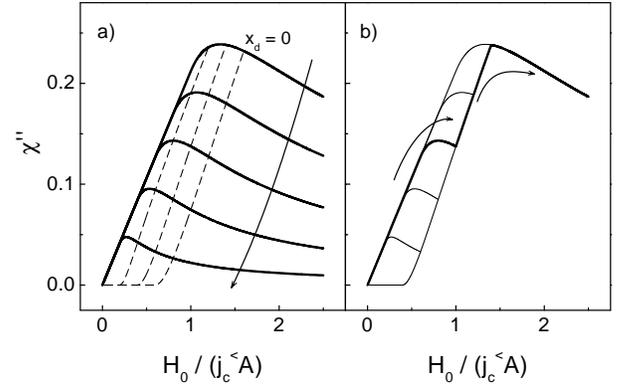}
\vspace{-20mm}

\caption{Dissipative component of the ac susceptibility, $\chi''$,
following Eq. \ref{e:chi1} and \ref{e:chi2}. (a)  $\chi''$ vs.
$\frac{H_{0}}{H_p}$ for $x_d/A=$ 0, 0.2, 0.4, 0.6, 0.8 (solid thick-lines). 
The arrow indicates increasing $x_d/A$. $x_d=0$
corresponds to the usual critical state. Dotted lines represent
the curves of minimal penetration defined in the text for
$h_{T}=$ 0.2, 0.4 and $0.6 H_p$.  (b) Susceptibility for $h_{T}=0.4
H_p$ and initial $x_d=x_{d}^i/A=0.4$ (solid thick-line). The arrows
indicate the direction of the magnetic field variation.}
\label{f:susslab}
\end{figure}

As an example, we plot in Fig. \ref{f:susslab}(a) the calculated
imaginary component $\chi''$ as a function of the normalized ac
magnetic field $H_0 /H_p$ for different values of $x_d/A$. These
results describe the case of a fixed boundary separating high and
low mobility regions of the VL (in each curve the boundary is at a different
position). We will address now the more realistic situation where
the ac field modifies the position of the boundary.

As was discussed above, the elastic pinning regime is valid for ac
fields of low amplitude, where vortex displacements at the sample
boundary do not exceed the size of the pinning potential well. As
the ac field amplitude is increased, the region in the sample
where this condition is valid will be gradually reduced. As a
consequence, the curves plotted in Fig. \ref{f:susslab}(a) for
fixed $x_d$ can not be valid for an arbitrary high ac field
amplitude. A more realistic approach has to consider that $x_d$
will gradually reduce for sufficiently strong fields. This
reduction should be monotonic in the ac field amplitude.

It is useful to define a ``minimal penetration susceptibility"
that represents the ac susceptibility of a sample that initially
contains a homogeneous low mobility VL and is subjected to an
increasing ac magnetic field starting from zero. These curves are
represented in Fig. \ref{f:susslab}(a) by dotted lines and will be
discussed in the following paragraphs.

The application (in the elastic regime) of a small perturbation
field $b_s$ (ac or dc) induces currents that exert a pressure on
the vortices at the sample boundary that propagates inside the
sample due to the VL elasticity. In a wide slab, the perturbation
decays exponentially, with a characteristic length $\lambda_C$,
the Campbell penetration depth.\cite{campbell69,
campbell71,campbell72,brandt92}

We suppose first an increasing ac field amplitude starting from
$H_0 =0$. As $H_0$ is increased above a given threshold,
$h_{T}$, vortices in the outer part of the sample are driven out
of the Campbell state as they are forced to move out of their
pinning sites. The subsequent oscillatory motion of the vortices
produces a reordering of the VL that results in a local reduction
of $j_c$ to $j_c^<$. Here, $h_{T}$ is related to the depth of the
pinning potential well in which vortices oscillate in the
Campbell regime.

We can estimate $x_d$ as a function of $H_0$ for a semi-infinite
sample ($x \geq 0$). A perturbation $b_s$ penetrates the sample
as $\Delta B(x) = b_s \exp(-x/\lambda_{ac}) $, where the ac
penetration depth is given by $ \lambda_{ac}^2 = \lambda^2 +
\lambda_{C}^2$ with $\lambda$ the London penetration
depth.\cite{brandt91,brandt92} In the case we are interested in
$\lambda_{C} = (c_{11} /\alpha_L)^ {1/2}$, where $ c_{11}$ is the
compressibility of the VL and $\alpha_L$ is the Labusch parameter
(in general for weak pinning $\lambda_{C}\gg \lambda$ and
$\lambda_{ac} \sim \lambda_{C}$).

To determine whether the elastic limit is applicable or not, we
estimate the vortex displacement at the sample surface as

\begin{equation} \label{e:desp}
\Delta x \sim \int_{0}^{\infty}\frac{\Delta B(x)}{B_{dc}}dx
\end{equation}

As discussed above, for displacements $\Delta x \sim r_p$, or
larger, the elastic regime is not expected to be applicable. We
find from Eq. \ref{e:desp} that this occurs for an applied
perturbation $b_s = b_{T} \sim r_p \sqrt{\alpha_L \mu_0}$, where
we have used $\lambda_{ac} \sim \lambda_{C}$ and $ c_{11} \sim
B_{dc}^2/ \mu_0$ ($b_T=\mu_0 h_T$). If the VL is not strongly
perturbed then $\alpha_L r_p = j_c^> B_{dc}$,\cite{brandt86}
leading to $b_{T} \propto \sqrt{j_c^>}$.

In the semi-infinite sample, it is evident that this relation
establishes the applicability of the elastic Campbell regime at
any point inside the sample. Then, we can assume that, if the VL
is in a low mobility state ($j_c =j_c^>$) in the whole sample, and
the ac field amplitude is increased from zero, $x_d$ will reduce
monotonically from $x_d = A$ to a value given by $\Delta B(x_d) =
b_{T} \propto \sqrt{j_c^>}$. In this picture, the ac
susceptibility of a slab is given by Eqs. (1)-(4) with $x_d$ not a
constant anymore but depending on the applied ac field amplitude,

\begin{eqnarray}
H_{0}\leq h_{T}:& x_d=A \nonumber \\ H_{0} > h_{T}: &
x_d=A-\frac{H_{0}-h_{T}}{j_c^{<}} \nonumber
\end{eqnarray}

Fig. \ref{f:susslab}(a) shows in dotted lines the ``minimal
penetration curve" as defined above for the imaginary component
$\chi''$ as a function of the normalized ac magnetic field $H_0
/H_p$ for $h_{T}$ = 0.2, 0.4 and 0.6 $H_p$. In Fig.
\ref{f:susslab}(b), we represent the ac susceptibility of a slab
predicted by this simplified picture where the initial state
corresponds to a situation in which the vortices in the outer
region, $|x| > x_{d}$, have a higher mobility ($j_c=j_c^<$). In
this example, we choose $h_{T} = 0.4 H_p$ and we show $\chi''$ for
different initial values of $x_{d}$ ($x_{d}^i$). The curve represented with a thick line
corresponds to $x_{d}^i= 0.4 A$. For $H_0 < H_d$ the
internal region of the sample provides no contribution to
$\chi''$, and the susceptibility corresponds to a slab in critical
state. For $H_d < H_0 < H_d + h_{T}$ (in this example $H_d +
h_{T}= H_p$) $\chi_{ac}$ moves away from the CS susceptibility as
the internal region contributes with a higher screening capability
and the dissipation decreases. However, as $\Delta B(x_d) <
b_{T}$, $x_{d}$ remains unchanged and equal to its initial value,
$x_{d}^i$. For $H_0 > H_d+h_{T}(=H_p)$ the boundary between low
and high VL mobility progressively penetrates into the slab and
$\chi_{ac}$ follows the minimal penetration curve. Finally, for
$H_0=H_p+h_{T}(=1.4 H_p)$ the sample is fully penetrated and the
CS regime is recovered. The convenience of this calculation will
become clear when we describe in Section IV the protocols followed
in experiments.

\subsection{\label{dyn}Numerical calculations}

\subsubsection{Computational method}

The original Bean critical state model has been extended to
different geometries, including long
strips\cite{brandt93,zeldov94} and circular
disks\cite{zhu93,mikheenko93,clem94} in a perpendicular magnetic
field. These geometries are more appropriate for our experimental
situation. The results obtained analytically can be reproduced by
an algorithm developed by E. H. Brandt.\cite{brandt94a,brandt94b}
The great advantage of the numerical approach is that it allows
to analyze static and dynamic problems in samples where
the critical current density has a non-trivial dependence on
position. The superconductor is modeled by a constitutive law
$E(j)$ that has the form $E(j) = E_0 (\frac{j}{j_c})^p$
corresponding to a logarithmic activation energy $U(j) \sim \ln
(j_c/j)$ or to $U(j) \sim (j_c/j)^{\alpha}$ with $\alpha\ll 1$ for
a superconductor with critical current density $j_c$.

Following Brandt, we have considered a strip localized in $| x |
\leq d/2\ll A $ with $| y |\leq A$ and $| z | \gg A$ in an applied
field $H_a \mathbf{\hat{x}}$ and we define the sheet current
density $\mathbf{J}(y,z)= \int_{-d/2}^{d/2} \mathbf{j}(x,y,z) dx$,
which in this case is $\mathbf{J}(y,z)=-J(y) \mathbf{\hat{z}}$
(the extension of these calculations for disks is straightforward
and is described in Ref. \onlinecite{brandt94b}).

The magnetic field generated by these currents is given by
Amp\`ere's law. For the strip, the total field perpendicular to
its surface evaluated at $x=0$ has the form

\begin{equation}
H_{x}(0,y)=H(y)=\frac{1}{2\pi}\int_{-A}^{A}\frac{J(u)}{y-u}du +
H_a \label{e:camp}
\end{equation}

As the current flowing in the strip is induced by the external
magnetic field, the symmetry  $J(y)=-J(-y)$ can be used in Eq.
\ref{e:camp}. Considering the electric field
$E(y,t)=\dot{\phi}(y,t)/L$, induced by the time dependent magnetic
flux $\phi(y)=\mu_0\int_0^y H(u)du$, one gets an
integro-differential equation for the sheet current. This equation
has been discretized\cite{brandt94a,brandt94b}

\begin{equation}
\dot{J_i}(t)=\sum_{j=1}^N K_{ij}^{-1}\left[\frac{J_j(t)}{\tau}-2
\pi
 u_j \dot{H}_a(t)\right]\label{e:discret}
\end{equation}

\noindent where $\tau = \mu_0 ad/(2 \pi \rho)$, $\rho=E(j)/j$ and
$N$ is the number of points $u_i$ $(0<u_i<1)$ of the spatial
grid, $u_i=u(x_i)$ with $x_i=(i-\frac{1}{2})/N$. In Eq.
\ref{e:discret} the strip half width, $A$, is the unit of length
and $K_{ij}$ is given by (see Ref. \onlinecite{brandt94a})

\begin{eqnarray}
K_{ij}=& \frac{w_j}{N}\ln \left|\frac{u_i-u_j}{u_i+u_j}\right| &
i\neq 0 \nonumber \\&\label{e:kij}\\
K_{ij}=& \frac{w_j}{N}\ln \frac{w_j}{4\pi Nu_i} & i=j \nonumber
\end{eqnarray}

\noindent where $w_i=u_i'$ is the weight function.

Having obtained the sheet current as a function of time by solving
Eqs. (\ref{e:discret})-(\ref{e:kij}), we evaluate the magnetic moment
$\mathrm{\mathbf{m}}$ of the sample that points in the
$\mathbf{\hat{x}}$ direction as

\begin{equation}
\mathbf{m}=\frac{1}{2}\int \mathbf{j}\times \mathbf{r}\: d^3x
\end{equation}

\noindent which for the strip results in

\begin{eqnarray}
\mathbf{m}=-\mathbf{\hat{x}}ML, &M=\int_{-A}^{A}yJ(y)dy
\end{eqnarray}

\noindent If the applied field is sinusoidal,
$H_a(t)=H_{ac}(t)=H_{0}\cos (\omega t)$ where $\omega=2\pi/T$, it
is possible to define the ac susceptibility coefficients as

\begin{eqnarray}
\chi'_n=\frac{\omega}{\pi H_{0}} \int_0^T M(t)\cos(n\omega t)\:
dt\\ \chi''_n=\frac{\omega}{\pi H_{0}} \int_0^T M(t)\sin(n\omega
t)\: dt
\end{eqnarray}

To obtain the susceptibility we solved Eqs.
(\ref{e:discret})-(\ref{e:kij}) by a fourth order Runge-Kutta method
with automatic control of the step size.\cite{NumRec} We chose
the constitutive relation suggested above,

\begin{equation}
 \mathbf{E}(\mathbf{J})=E_c
\left(\frac{J}{J_c}\right)^p \frac{\mathbf{J}}{J} \label{e:relc}
\end{equation}

\noindent with $J=|\mathbf{J}|$.

In the Bean critical state model  $E(J)$ is strongly nonlinear and
has an abrupt increment at $J=J_c$. For the constitutive relation
of Eq. \ref{e:relc} this occurs for $p \rightarrow \infty$. For $1
\ll p < \infty$, it leads to a nonlinear relaxation (approximately
logarithmic) of $M(t)$  caused by the nonlinear magnetic flux
diffusion. As $p$ decreases the relaxation becomes stronger and in
the limit $p=1$ the behavior is ohmic, with linear magnetic flux
diffusion and exponential relaxation of $M(t)$. In our simulations
we chose $p=20$, as $p<10$ would blur the flux front due to flux
creep and $p>50$ would require long simulation times to detect
relaxation effects. Note that, in general, only the combination
$E_c/J_c^p$ is relevant (Eq. \ref{e:relc}), and the voltage
criterion, $E_c$, and $J_c$ are not independent. While in the
critical state limit $E_c$ is irrelevant, in the ohmic  regime
$E_c/J_c=\rho$  is the resistivity. Within the dynamical approach
that we describe, $J_c$  has a different meaning. As the current
is time dependent (it may depend on the rate of variation of the
magnetic field), $J_c$ is related to the degree of vortex
mobility, where low values of $J_c$ represent highly mobile vortex
structures.

The space variable was discretized with $N=50$ nonequidistant
points (results with $N=30$ showed no significant difference). The
tabulation was performed so that the weight function $w(x)=u'(x)$
vanished at the integration boundary to remove the infinity in the
integrand occurring at $u \rightarrow 1$ at $t= 0$ (see Ref.
\onlinecite{brandt94a})

\begin{eqnarray}
u(x)&=&\frac{3}{2}x-\frac{1}{2}x^3 \nonumber\\
w(x)&=&\frac{3}{2}(1-x^2) \nonumber
\end{eqnarray}

For the ac field $H_{ac}(t)=H_{0} \mathrm{cos}(\omega t)$ measured
in units of $j_c d$, $\omega=2E_c/(\mu_0$ $j_cdA)$ which, in units
of $A=2E_c=j_cd=1$ is $\omega=1/\mu_0$.  The ac susceptibility was
calculated after two complete ac cycles that are necessary to
reach a stationary state in the calculation of the hysteresis loop
$M[H(t)]$.

\subsubsection{Current and field profiles. ac susceptibility}

\begin{figure*}


\includegraphics[width=7in]{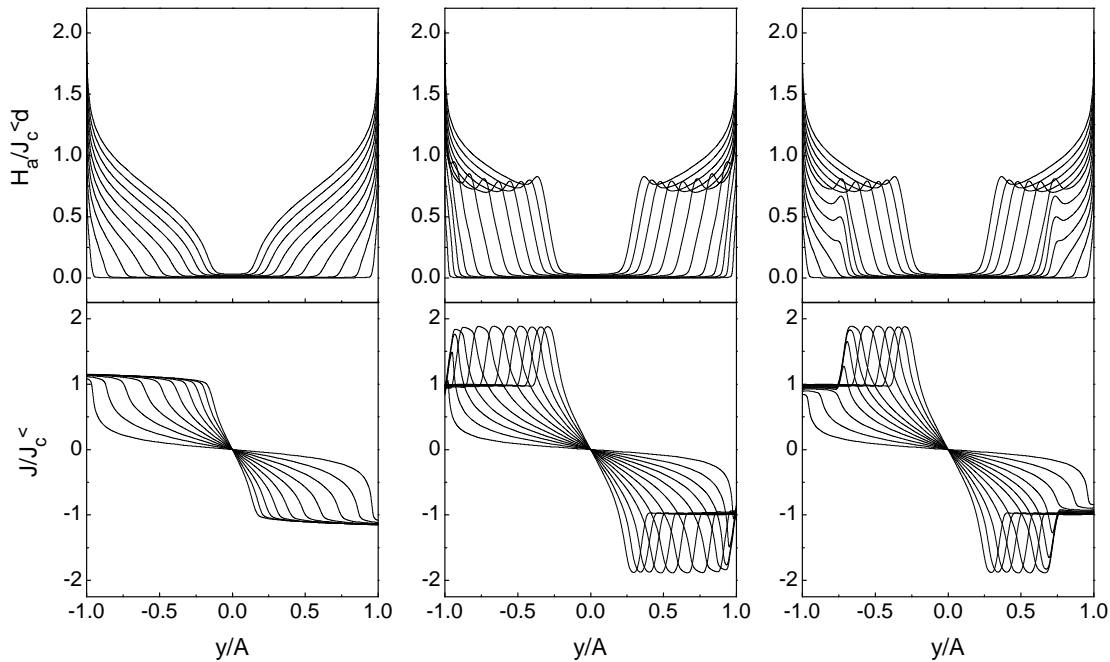}
\vspace{-25 mm}

\caption{Field profiles (top) and current profiles  (bottom)
obtained numerically for a field $H_0\cos \omega t$ increasing
from zero and evaluated at $H_0$, for different $H_0<1$. Left:
homogeneous sample, $J_c^>=J_c^<=J_c$. Center: penetration for the
minimal penetration curve $J_c^>=2J_c^<$. Right: profiles obtained
increasing $H_0$ with initial $x_{d}^i=0.75A$. Calculations are
for a thin superconducting strip. See the text.}
\label{f:perfiles}
\end{figure*}

\begin{figure*}
\includegraphics[width=7in]{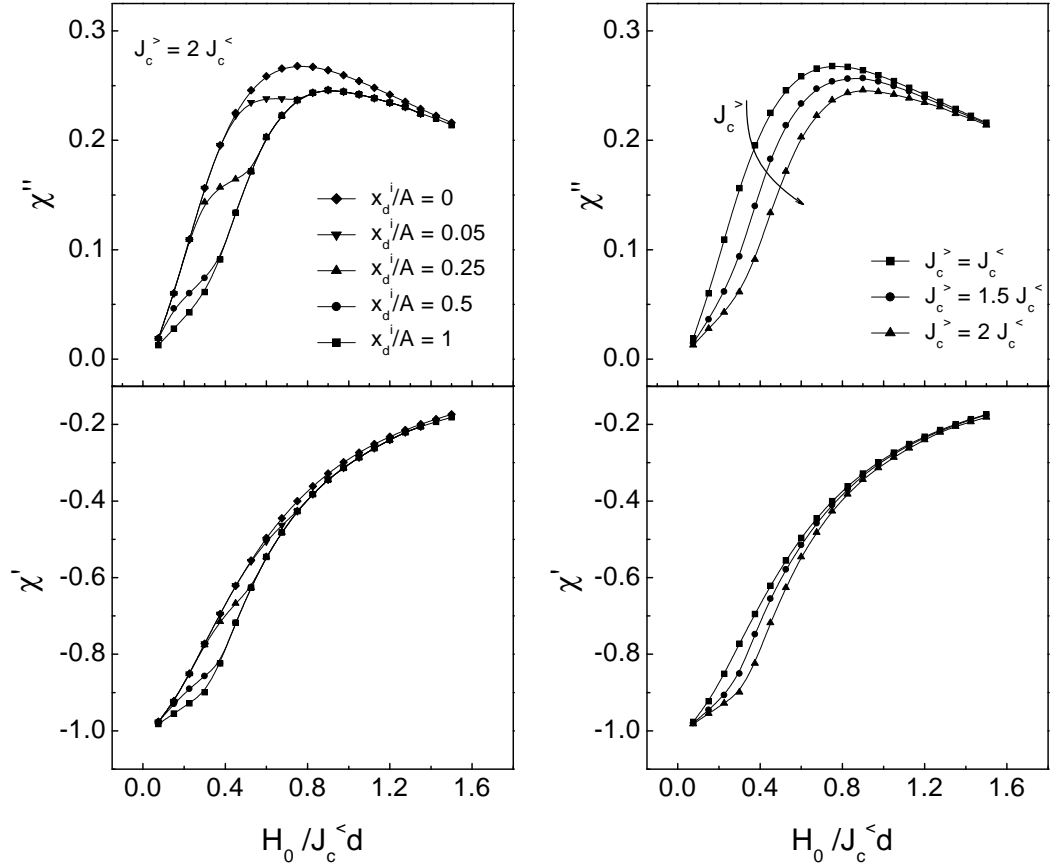}
\vspace{-10mm}
\caption{Numerical calculation of the ac susceptibility of a
strip, $\chi''$ (top) and $\chi'$ (bottom). On the left, we show
results increasing the field for different initial values of
$x_d$, $x_{d}^i/A=$ 0, 0.05, 0.25, 0.5 and 1. $J_c^> =2J_c^<$. On
the right, we show the minimal penetration curves for  $J_c^>=$ 1,
1.5 and 2 $J_c^<$. They are obtained by taking $x_d^i$ equal to
zero. See the text.} \label{f:renum}
\end{figure*}

In our calculations, the critical sheet current $J_c$ at each
point $y_i$ inside the strip depends on the magnetic history of
the sample. As was discussed above, we assume that $J_c$ can take
two different values, $J_c^>$ or $J_c^<$ (now both finite)
depending on the sheet current $J(y_i)$ that flowed in $y_i$ in
the previous iteration. For a given $H_0$, if $J(y_i) > J_c^>$
then $J_c(y_i)$ is replaced by $J_c^<$ (which has been
arbitrarily chosen as unity) and the calculation is restarted.
Clearly, the magnetic history of the sample fixes  the initial
distribution of $J_c(y)$. In  Figs. \ref{f:perfiles} and
\ref{f:renum} we show the result of our calculations starting with
zero ac field and different initial conditions. In Fig.
\ref{f:perfiles} we show the field (upper panel) and the current
(lower panel) distributions for a strip with the following initial
critical sheet current distributions:  $J_c$ =$J_c^<$ (left),
$J_c$ =$J_c^>$ (center), and $J_c$ =$J_c^<$ for $| y/A | \geq
0.75$ and $J_c$=$J_c^>$ for $| y/A | \leq 0.75$ (right panels).

In the protocol for the calculation of ac susceptibility (Fig.
\ref{f:perfiles}) we  sweep the ac field from zero. The minimal
penetration curve is calculated as follows. We begin with
$J_c=J_c^>$ for the whole strip (low mobility initial state,
center panels in Fig. \ref{f:perfiles}). The field is increased
from zero and we compute the field and current profiles. If the
sheet current exceeds $J_c^>$ at any point of the strip, then
$J_c$ is changed to $J_c^<$ at that point and the calculations are
restarted. This generates by construction the smallest outer
region with $J_c=J_c^<$ for each given ac amplitude $H_0$ (so
that $J<J_c^>$ in the whole sample). The field amplitude $H_0$ is
increased again and the procedure repeated. Note that the region
with $J_c=J_c^>$ shrinks towards the center of the strip as $H_0$
is increased. In the calculation of $\chi_{ac}$ the field has to
be varied in a complete cycle, but it is clear that the critical
current density distribution is fixed after the first quarter of
cycle.

Now we calculate the rest of the curves that result for a strip
with an initial distribution of $J_c$ as in the right panels of
Fig. \ref{f:perfiles}. To obtain the initial state, we apply a
magnetic field of a given amplitude and we define the initial
distribution of $J_c$ following the procedure described above.  We
then calculate the ac susceptibility increasing $H_0$ from zero
and check for the condition to replace $J_c^>$ by $J_c^<$. This
occurs for field amplitudes larger than the one that defined the
initial state. Note that from this field amplitude on, the
calculated susceptibility coincides with the minimal penetration
curve.

It is worth noting that in a strictly 2D model, it is not possible
to have an abrupt change in $J_c$, because this causes a local
divergence in $J$ and vortices would always move.  To solve this
inconvenience, in our calculations, the critical current density
changes smoothly from $J_c^>$ to $J_c^<$ over distances of the
order of $d$ (where the 2D treatment breaks down\cite{clem94}).

In Fig. \ref{f:renum} (left panels) we took $J_c^>=2J_c^<$. In
Fig. \ref{f:renum} (right panels) we compare the minimal
penetration curves for different values of $J_c^>$ (1, 1.5 and 2).
As we shall see in the next section, the protocol followed in the
numerical calculations is particularly convenient to directly
compare the numerical results with the experimental ones. In the experiments
that are described below, we are able to prepare, in a highly
reproducible way, an initial state with low mobility in the whole
sample (that is, described in calculations with $J_c^>$) or an
initial state with high mobility at the outer region of the sample
(a situation that corresponds to $J_c^<$ in the outer region and
$J_c^>$ in the inner region).

\section{\label{exp}Experiment}

We measured the ac susceptibility of a twinned
YBCO single crystal \cite{alexandrov88}
using a standard mutual inductance technique. The characteristics
of the crystal are: dimensions $0.56 \times 0.6 \times 0.02 ~
\mathrm{mm}^{3}$,  $T_{\mathrm{c}}$ = 92 K and $\delta T_c$ = 0.3
K determined by ac susceptibility ($h_{ac}$=1 Oe) at zero dc
field. In our measurements the ac field is applied parallel to the
\textit{c} axis of the sample. The dc magnetic field $H_{dc} =
3$ kOe was applied at $20^{\circ}$ out of the twin boundaries to
avoid the Bose-glass phase. Further details can be found in Ref.
\onlinecite{SOVPRL00}.

The generally unknown dependence of the critical current on
temperature makes $\chi_{ac}$ measurements vs. ac field amplitude
at constant temperature  the best choice for comparison with
numerical results. The experimental procedure, based on our
previous results, is as follows:

\begin{itemize}
\item We set the VL in a low mobilty state by applying $10^4$
cycles of an ac magnetic field with a sawtooth waveform (10 kHz, 8
Oe, see Ref. \onlinecite{SOVPRL01}). This field completely
penetrates the sample.

\item We define the initial state by applying $10^5$ cycles of a
sinusoidal ac field of a given amplitude at 10 kHz.

\item This ac field is turned off and we measure ac susceptibility increasing the amplitude of the
ac field probe from zero, with a fixed measuring frequency of 10
kHz.
\end{itemize}

\begin{figure}[t]
\vspace{-10mm}
\includegraphics[width=4.5in]{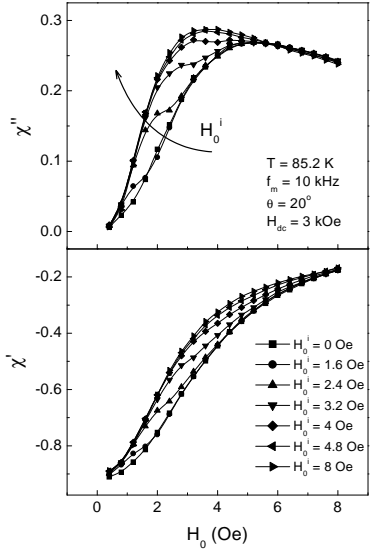} \vspace{-8mm}
\caption{Experimental results. $\chi'$ and $\chi''$ for a YBCO
single crystal as a function of the ac field amplitude, $H_0$. The
measurements were performed increasing $H_0$ from zero. The
different curves correspond to different initial conditions.
Before the measurement starts, a low mobility VL is prepared in
the whole sample and an ac field of amplitude $H_0^i$ and
frequency $f=10$ kHz is applied to define the initial state.
$H_0^i=$ 0, 1.6, 2.4, 3.2, 4, 4.8, and 8 Oe.}
\label{f:susord}
\end{figure}

Results are shown in Fig. \ref{f:susord} for a fixed temperature
and $H_{dc}=3$ kOe. Different initial states were prepared
applying $10^5$ cycles of a sinusoidal ac field of amplitudes  0,
1.6, 2.4, 3.2, 4, 4.8 and 8 Oe.

The similarities with numerical simulations for $J_c^>=2J_c^<$
shown in Fig. \ref{f:renum} are evident. The marked change in the
response at the ac field amplitude that defines the initial state
supports the scenario of vortices being shaken or not out of the
pinning centers on either side of a well-defined boundary. The
effective critical current density of the more mobile state can be
estimated from the position of the dissipation peak. Comparing
Figs. \ref{f:renum} and \ref{f:susord} we have $j_c^<\simeq
H_{0}^{peak}/(0.75d)$. With $d=30 \:\mu$m and $H_{0}^{peak}=3.5$
Oe, $j_c^<\simeq 1250$ A/cm$^2$, and thus $j_c^>\simeq 2500 $
A/cm$^2$ (at 85.2 K).

We also observe that the inductive component does not approach -1
as $H_{0}$ is reduced to zero. This is probably a consequence of
the finite penetration depth in the Campbell
regime.\cite{brandt94c,pasquini} We can determine some of the
relevant parameters in this regime, estimating from the
saturation value ($\chi'(0) \sim$ -0.9), the value for
$\lambda_{C} \sim \lambda_{ac} \sim 10 \: \mu$m ($\gg \lambda
\sim 0.6\:\mu$m, Ref. \onlinecite{poole}) and the size of the
potential well $r_p \sim 50\:\mathrm{ {\AA}}$ (Eq. \ref{e:desp} $r_p
\sim \lambda_{ac} h_{T}/H_{dc}$ with $h_{T} \sim 1.5$ Oe and
$H_{dc}=3$ kOe). The value for $h_{T}$ was estimated as the
minimum ac amplitude capable of modifying the measured
susceptibility (starting from a low mobility VL).

We note that from these estimates $r_p$ turns out to be of the
order of the coherence length  $\xi \sim 60\:\mathrm{ {\AA}}$ at
our working temperature ($\xi \sim \xi_0 (1-T/Tc)^{-1/2}$ with
$\xi_0\sim 20\:\mathrm{ {\AA}}$, Ref. \onlinecite{poole}). The
Labusch parameter is $\alpha_L \sim \frac{j_cB_{dc}}{r_a} \sim
\frac{j_c\phi_{0}}{r_a}\frac{1}{a_0^2}\sim 10$ N/m$^2
\frac{1}{a_0^2}$ (100 dyn/cm$^2 \frac{1}{a_0^2}$).

\subsubsection{Frequency dependence of the penetration depth}

Within the dynamical model described above, an increase in the
frequency of the ac field induces a higher ac current (on the
average) with a smaller penetration depth. One can gain some
insight by considering a stationary state where a constant field
ramp $\dot{H}_a$ is applied. In the stationary state the current
induced by the field ramp does not change with time and the
electric field is obtained as
$\nabla\times\mathbf{\mathrm{E}}=-\mathrm{\mathbf{\dot{B}}}_a$,
where $\mathrm{\mathbf{B}}_a=\mu_0\mathrm{\mathbf{H}}_a$, implying
$E \sim \dot{B}_a$. Inverting Eq. \ref{e:relc} we get $J\sim
(\dot{B}_a)^{1/p}$, which shows how $J$ grows as $\dot{B}_a$ is
increased. Taking $J$=constant in, for example, Eq.
\ref{e:discret}, we arrive at the same result.

The reduction in penetration due to an increment in the frequency
can be deduced using an interesting scaling property of Eq.
\ref{e:discret} (Ref. \onlinecite{brandt97}). Explicitly, if the unit of time is
changed by a factor $k$ and time is expressed as $\tilde{t}=t/k$,
then the new functions $\tilde{J}(\tilde{t})=J(t)\:k^{1/p-1}$ and
$\tilde{B_a}(\tilde{t})=B_a(t)\:k^{1/p-1}$ satisfy the same
equations. The resulting magnetic field scales with the same
factor. If we consider a periodic magnetic field,
$H_{ac}(t)=H_0\mathrm{sin}\omega t$, this scaling property implies
that if the frequency is increased by a factor 10 and the ac field
amplitude is also increased but by a factor 10$^{(1/p-1)}$, the
field penetration depth inside the sample will remain
unchanged.\cite{brandt97} This leads to the immediate conclusion
that if the frequency is increased but the amplitude of the field
is maintained constant, then the penetration depth will be
smaller.

\begin{figure}[t]
\vspace{-10mm}
\includegraphics[width=4.5in]{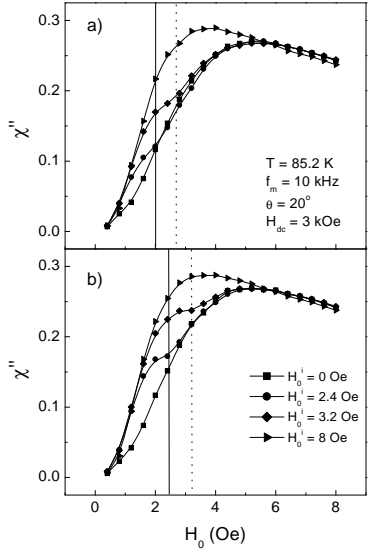} \vspace{-8mm}
\caption{Experimental results analogous to those shown in Fig.
\ref{f:susord}. To define the initial state, we applied an ac
field with the amplitudes used in Fig. \ref{f:susord} but with
frequencies 10 times higher ($f=100$ kHz)[(a)]. As a comparison we
have included results for $\chi''$ from Fig. \ref{f:susord} in
panel (b). It is clear that in the latter case the field that
defines the initial state (10 kHz) penetrates deeper into the
sample. This is because the boundary between low and high critical
current densities is found at a higher $H_0$ when the initial
state is defined with 10 kHz as noted by the vertical lines. In
other words, the vertical lines compare the corresponding ac field
amplitudes for which the measuring field has penetrated inside the
sample a length similar to that penetrated by the field that
defines the initial state.} \label{f:susord-}
\end{figure}

This behavior is observed experimentally. In Fig. \ref{f:susord-}
we compare the imaginary part of the susceptibility measured at 10
kHz after applying 10$^5$ cycles of a sinusoidal ac field
of a given amplitude at 100 kHz [Fig. \ref{f:susord-}(a)] and at 10
kHz [Fig. \ref{f:susord-}(b)]. As the amplitude of the measuring
field increases, the susceptibility senses the VL mobility in the
regions closer to the center of the sample. As the change in the
response towards the low dissipation branch is first observed 
in Fig. \ref{f:susord-}(a), it is clear
that the penetration at 100 kHz is smaller than at 10 kHz. The reason for this is that, 
in the former case, the high-mobility outer-region is necessarily smaller 
as implied by the susceptibility measurements (note
that the susceptibility is measured at 10 kHz). From this
difference, and applying the scaling law discussed above, we can estimate the
exponent $p$ in Eq. \ref{e:relc}. As an example, the penetration
depth of 3.2 Oe at 100 kHz is approximately equal to the
penetration of 2.7 Oe at 10 kHz. Assuming a linear dependence of
the penetration depth with the ac field amplitude, we obtain $p \sim 15$,
which is close to the value we used in our
simulations.

\subsubsection{Temporarily symmetric and asymmetric ac magnetic fields}

We have suggested in Ref. \onlinecite{SOVPRL01} that a change in
the mobility of the VL may be related to a change in its degree of
order, and that these changes are intrinsic to the oscillatory
dynamics. Molecular dynamics calculations\cite{SOVPRL02} support
the scenario where defects are healed by a temporarily symmetric
oscillation and created by a temporarily asymmetric oscillation.
These simulations also show that the mean velocity of the VL
increases as the number of defects decrease.

In the previous section we showed that it is possible to increase
vortex mobility (or to order the VL) by applying a temporarily
symmetric ac field, and we proved that this reordering occurs in
the outer penetrated region of the sample, {\it i.e.} in the
regions where induced currents have densities above the critical
current density. We also know that a temporarily asymmetric field
produces the opposite effect: it leads the VL into a low mobility
(disordered) state.\cite{SOVPRL01} Then, the following questions
arise: \textit{i)} Does the penetration depth of the disordered
region have a similar dependence on ac amplitude as the
penetration depth of the ordered region for symmetrical ac
magnetic fields?
\textit{ii)} Does the amplitude of the asymmetric field have to
be above a certain threshold to produce any effect?

Related to this last question we found that, for the application
of a symmetrical field, the minimum amplitude that produces some
reordering is $\sim 1.5$ Oe, and this value may be related to the
depinning of vortices out of their pinning sites. As we shall see, however, the threshold amplitude for the
asymmetric field to produce some disordering is twice as high, thus
the process that prevents the VL from switching to a more pinned state should have a different origin.

The experimental protocol for the asymmetric ac fields is very
similar to the one described above for symmetric fields, but was
modified so that the mobility of the initial state is low. The
following procedure is used:

\begin{itemize}
\item We generate a high mobility VL in the whole sample by
exposing it to  $10^4$ cycles of an oscillating magnetic field
with a sawtooth waveform (8 Oe amplitude and 10 kHz) followed by
$10^5$ cycles of a sinusoidal ac field of the same amplitude and
frequency. This leads to a high mobility VL in the whole sample.

\item We define the initial state by applying $10^5$
\textit{asymmetric} ac magnetic field cycles (\textit{i.e.},
sawtooth) of a given amplitude, $H_0^i$ (10 kHz).

\item We measure ac susceptibility with a sinusoidal ac probe as a
function of amplitude, starting from amplitude zero (measuring
frequency: 10 kHz).

\end{itemize}

\begin{figure}[t]
\vspace{-10mm}
\includegraphics[width=4.5in]{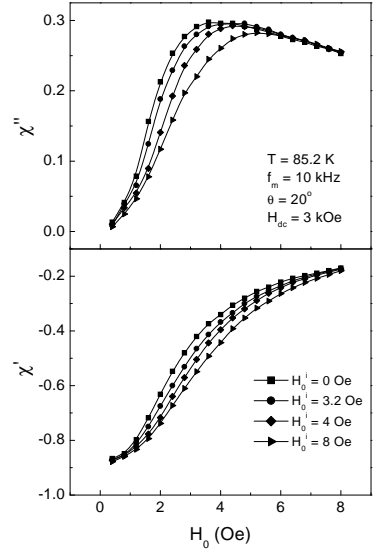} \vspace{-8mm}
\caption{$\chi'$ and $\chi''$ in increasing field for different
initial conditions. The mobility of the VL is reduced when
asymmetric ac fields are applied. Before the measurement starts, a
high mobility VL is prepared in the whole sample and a sawtooth
field of amplitude $H_0^i$ and frequency $f=10$ kHz is applied to
define the initial state. $H_0^i = $ 0, 1.6, 4, and 8 Oe.} \label{f:susdes}
\end{figure}

Results are plotted in Fig. \ref{f:susdes} for the same
temperature and field conditions of Figs. \ref{f:susord} and
\ref{f:susord-}. We investigated different initial states prepared
by the application of a sawtooth waveform ac field with amplitude,
$H_0^i$, between 0 and 8 Oe (we show the results for 
$H_0^i$ = 0, 3.2, 4, and 8 Oe). The first observations
are that there seems not to be a clear boundary separating two
regions with different vortex mobilities as in Fig. \ref{f:susord},
and that changes in susceptibility are important only for $H_0^i \sim 3$
Oe or higher. With asymmetric ac fields, the curves for the different $H_0^i $ resemble the minimal
penetration curves of Fig. \ref{f:renum} for different
$J_c^{int}$s. The fact that there is no clear boundary between
two regions with associated different mobilities of the VL might
be explained by the difference in the sweep rate of the sawtooth
waveform that implies a different penetration depth in the ramp up
and down of the field. The measuring ac field can also contribute
to blur the boundary as it tends to counteract the effect of the
sawtooth ac field. On the other hand, we cannot obviate the fact
that the shape of the susceptibility does not favor its
observation. Finally, another possible explanation for this
behavior is related to the mechanism that leads to the decrease of
the VL mobility. In Ref. \onlinecite{SOVPRL01}, we proposed that
this decrease is related to an increase in the density of
topological defects due to plastic distortions in the VL
that are produced by the temporarily asymmetric induced currents 
(cf. Fig. \ref{f:susord-}, it is implicitly shown that higher field rates induce higher currents). The
observation of a minimum amplitude $\sim 3$ Oe required to
produce a strong effect in the mobility of a moving
lattice supports this assumption. For plastic distortions to occur, we
expect that vortices should move around one lattice parameter.
For a sample in the critical state, the displacement at the sample
boundary caused by a field that penetrates a distance $d$ is
$\Delta x\approx \frac{1}{4}\frac{H_0}{B_{dc}}d$ (for $d$ smaller
than the sample radius). For a sample of radius $250$ $\mu$m in a
3 kOe dc field and 3 Oe ac field (6 Oe peak to peak, and 5
Oe of full penetration) the total vortex displacement at the
sample periphery is $\sim 0.1$ $\mu$m, which is of the order of
the lattice parameter $a_0 = 0.09$ $\mu$m.

We note that although an important portion of the VL is moving by
the application of 3 Oe ac field, only the portion near the sample
boundary is initially distorted. This is not what occurs during VL
ordering, where all the vortices that move by the action of a
symmetrical ac field organize in a more ordered structure. It is
possible that the defects generated at the outer region of the
sample by the sawtooth waveform diffuse deeper into the sample
assisted by the moving lattice. As disorder is produced only for
high ac fields ($H_0 > 3$ Oe) that penetrate an important
fraction of the sample, a significant portion of the VL will be
contaminated after a certain number of oscillations. This description is supported by the direct
observation in molecular dynamics simulations\cite{SOVPRL02} of
the high mobility of dislocations in VL forced by oscillating
forces.

\section{\label{conc}Conclusions}

We have presented a phenomenological model that accounts for the
history effects in the the ac response of YBa$_2$Cu$_3$O$_{7}$
single crystals. The model is based on the effect of inhomogeneous
volume-pinning in the ac susceptibility. We obtained analytical
expressions for a simple geometry to address qualitatively the
details of the model, and then we presented numerical results by
solving Maxwell's equations for a more realistic situation.

We presented new experimental results that are adequate for
comparison with the numerical calculations. The very good
qualitative agreement between them indicates that we have captured
the essence of what occurs in relation to the different dynamical
responses of the VL inside the sample.

The existence of a threshold sawtooth ac magnetic field amplitude
required to lower the VL mobility supports the hypothesis of a
ratchet-tearing of the VL produced by temporarily asymmetric fields as
was suggested in Ref. \onlinecite{SOVPRL01}.

Some of the differences between experimental and numerical results
are probably  a consequence of an intrinsic property of the
experimental technique that undoubtedly perturbs the state of the
VL and in a way that may depend on the initial conditions.  A
sinusoidal ac field increases vortex mobility and the measurement
itself can mask the effect of a previously applied asymmetric ac
field. Besides, $J_c$ is not restricted to take only two values.
Future experimental and theoretical work is required to further
understand the oscillatory VL dynamics.

\begin{acknowledgments}
VB acknowledges financial support from CONICET. This research was partially supported by UBACyT TX-90, CONICET
PID N$^{\circ}$ 4634 and Fundaci\'on Sauber\'an.
\end{acknowledgments}


\begin{thebibliography}{42}
\bibitem{com} Note that, in the experiments of Ref.
\onlinecite{SOVPRL01}, the average amplitude of the vortex
oscillation is of the order of one lattice constant. In this
situation, the steady-motion theory of Ref.
\onlinecite{koshelev94} is not expected to apply.
\end{thebibliography}

\end{document}